\documentclass[twocolumn,aps,floats,superscriptaddress]{revtex4}
\usepackage{amsmath}
\usepackage{graphicx}

\newcommand{\mr}[1]{{{\mathrm{#1}}}}
\newcommand{\mcal}[1]{{\mathcal{#1}}}
\newcommand{\dt}{\partial_\tau}
\newcommand{\inte}{\int_0^\beta \!\!\!\! d\tau}
\newcommand{\w}{\omega}
\newcommand{\dsd}{d_{\sigma}^{\dagger}}
\newcommand{\ds}{d_{\sigma}}
\newcommand{\fsd}{f_{\sigma}^{\dagger}}
\newcommand{\fs}{f_{\sigma}}
\newcommand{\s}{\sigma}
\newcommand{\eps}{\epsilon_{0}}
\newcommand{\qbar}{\overline{Q}}
\def\ket#1{\left\vert #1 \right\rangle}

\newcommand{\double}{d_{\uparrow\downarrow}}
\newcommand{\myeq}{\!=\!}
\def\cN{{\cal N}}

\begin{document}

\bibliographystyle {plain}
\vspace{0.5cm}

\title{Quantum impurity solvers using a slave rotor representation}
\author{Serge Florens}
\author{Antoine Georges}
\affiliation{
Laboratoire de Physique Th\'eorique, Ecole Normale Sup\'erieure,
24 rue Lhomond, 75231 Paris Cedex 05, France }
\affiliation{ Laboratoire de Physique des Solides, Universit\'e Paris-Sud,
B\^{a}t.~510, 91405 Orsay, France}

\begin{abstract}
\vspace{1cm}
We introduce a representation of electron operators as a product of a spin-carrying
fermion and of a phase variable dual to the total charge (slave quantum rotor).
Based on this representation, a new method is proposed for solving
multi-orbital Anderson quantum impurity models at finite interaction strength $U$.
It consists in a set of coupled integral equations for the auxiliary field
Green's functions, which can be derived from a controlled saddle-point in the limit
of a large number of field components.
In contrast to some finite-$U$ extensions of the non-crossing approximation, the new
method provides a smooth interpolation between the atomic limit and the weak-coupling
limit, and does not display violation of causality at low-frequency.
We demonstrate that this impurity solver can be applied
in the context of Dynamical Mean-Field Theory, at or close to half-filling.
Good agreement with established results on the Mott transition is found, and
large values of the orbital degeneracy can be investigated at low computational
cost.

\end{abstract}

\maketitle

\tableofcontents

\section{Introduction}

The Anderson quantum impurity model (AIM) and its generalizations play a key role
in several recent developments in the field of strongly correlated electron systems.
In single-electron devices, it has been widely used as a simplified model for the
competition between the Coulomb blockade and the effect of tunnelling
\cite{quantum_dots}.  In a different
context, the Dynamical Mean-Field Theory (DMFT) of strongly correlated electron systems
replaces a spatially extended system by an Anderson impurity
model with a self-consistently determined bath of conduction electrons \cite{RMP_DMFT,pruschke-review}.
Naturally, the AIM is also essential to our understanding of local moment formation
in metals, and to that of heavy-fermion materials, particularly in the mixed valence regime
\cite{hewson}.

It is therefore important to have at our disposal quantitative tools allowing for the
calculation of physical quantities associated with the AIM.
The quantity of interest depends on the specific context. Many recent applications
require a calculation of the localized level Green's function (or spectral function),
and possibly of some two-particle correlation functions.

Many such ``impurity solvers'' have been developed over the years.
Broadly speaking, these methods fall in two categories: numerical algorithms which
attempt at a direct solution on the computer, and analytical approximation
schemes (which may also require some numerical implementation).
Among the former, the Quantum Monte Carlo approach (based on the Hirsch-Fye algorithm) and
the Numerical Renormalization Group play a prominent role. However, such methods become
increasingly costly as the complexity of the model increases.
In particular, the rapidly developing field of applications of DMFT to electronic structure
calculations \cite{LDA_DMFT} requires methods that can handle impurity models involving
orbital degeneracy. For example, materials involving f-electrons are a real challenge
to DMFT calculations when using Quantum Monte Carlo. Similarly, cluster extensions
of DMFT require to handle a large number of correlated local degrees of freedom.
The dimension of the Hilbert space grows exponentially with the number of these
local  degrees of freedom, and this is a severe limitation to all methods,
particularly exact diagonalization and numerical renormalization group.

The development of fast and accurate (although approximate) impurity solvers is therefore
essential. In the one-orbital case, the iterated perturbation theory (IPT) approximation
\cite{IPT} has played a key role in the development of DMFT, particularly in elucidating the
nature of the Mott transition \cite{IPT_Mott}.
A key reason for the success of this method is that it becomes exact both in the limit
of weak interactions, and in the atomic limit.
Unfortunately however, extensions of this approach to the multi-orbital case have not been
as successful \cite{kajueter_multi}.
Another widely employed method is the non-crossing approximation, or NCA \cite{NCA,U_NCA}.
This method takes the atomic limit as a starting point, and performs a self-consistent
resummation of the perturbation theory in the hybridization to the conduction bath.
It is thus intrinsically a strong-coupling approach, and indeed it is in the strong-coupling
regime that the NCA has been most successfully applied. The NCA does suffer from some
limitations however, which can become severe for some specific applications.
These limitations are of two kinds:
\begin{itemize}
\item The low-energy behaviour of NCA integral equations is well-known to display
non-Fermi liquid power laws. This can be better understood when formulating the NCA approach
in terms of slave bosons \cite{Cox_Ruckenstein,OP_AG_GK_AS}.
It becomes clear then that the NCA actually describes accurately the
overscreened regime of multichannel models. This is in a sense a remarkable success of NCA, but
also calls for some care when applying the NCA to Fermi-liquid systems in the screened regime.
It must be noted that recent progresses have been made to improve the low-energy behaviour
in the Fermi liquid case \cite{NCA_FLT}.

\item A more important limitation for practical applications has to do with the finite-$U$
extensions of NCA equations. Standard extensions do not reproduce correctly the non-interacting
$U=0$ limit. In contrast to IPT, they are not ``interpolative solvers'' between the weak-coupling
and the strong coupling limit. Furthermore, the physical self-energy tends to develop non-causal
behaviour at low-frequency ({\it i.e.} $\Sigma''_d(\omega)>0$) below some temperature (the ``NCA pathology'')
At half-filling and large $U$, this only happens at a rather low-energy scale, but away from half-filling
or for smaller $U$, this scale can become comparable to the bandwith, making the finite-$U$ NCA of limited
applicability.

\end{itemize}

In this article, we introduce new impurity solvers which overcome some of these difficulties.
Our method is based on a rather general representation of strongly correlated electron systems,
which has potential applications to lattice models as well \cite{hubbard_rotor}. The general idea is to introduce
a new slave-particle representation of physical electron operators, which emphasizes the {\it phase
variable dual to the total charge} on the impurity. This should be contrasted to slave-boson
approaches to a multi-orbital AIM: there, one introduces as many auxiliary bosons as there are
fermion states in the local Hilbert space. This is far from economical: when the local interaction
depends on the total charge only, it should be possible to identify a collective variable which
provides a minimal set of collective slave fields. We propose that the phase dual to the total
charge precisely plays this role. This turns a correlated electron model (at finite $U$)
into a model of spin- (and orbital-) carrying fermions coupled to a
{\it quantum rotor degree of freedom}. Various types of approximations can then be made
on this model. In this article, we emphasize an approximate treatment based on a sigma-model
representation of the rotor degree of freedom, which is then solved in the limit of a
large number of components. This results in coupled integral equations which share some
similarities to those of the NCA, but do provide the following improvements:
\begin{itemize}
\item The non-interacting ($U=0$) limit is reproduced exactly. For a fully symmetric
multi-orbital model at half-filling and in the low temperature range,
the atomic limit is also captured exactly, so that the
proposed impurity solver is an interpolative scheme between weak and strong coupling. The
sigma-model approximation does not treat as nicely the atomic limit far from half-filling
however (though improvements are possible, see section \ref{sec:doping}).
\item The physical self-energy does not display any violation of causality (even at low temperature
or small $U$). This is guaranteed by the fact that our integral equations become exact in
the formal limit of a large number of orbitals and components of the sigma-model field.
\end{itemize}
It should be emphasized, though, that the low-energy behaviour of our equations is similar to
the (infinite-$U$) NCA, and characterized by non-Fermi liquid power laws below some low-energy
scale.

As a testing ground for the new solver, we apply it in this paper to the DMFT treatment of the
Mott transition in the multi-orbital Hubbard model. We find an overall very good agreement with the
general aspects of this problem, as known from numerical work and from some recently
derived exact results. We also compare to other impurity solvers (IPT, Exact Diagonalization,
QMC, NCA), particularly regarding the one-electron spectral function.

This article is organized as follows.
In section \ref{sec:rotor}, we introduce a representation of fermion
operators in terms of the phase variable dual to the total charge
(taking the finite $U$, multi-orbital Anderson impurity model as an example).
In section \ref{sec:NCA}, a sigma-model representation of this phase variable
is introduced, along with a generalization from $O(2)$ to $O(2M)$. In the limit
of a large number of components, a set of coupled integral equations is derived.
In section \ref{sec:SIAM}, we test this ``dynamical slave rotor'' (DSR) approach
on the single-impurity Anderson model. The
rest of the paper (Sec.~\ref{sec:DMFT}) is devoted to applications of this approach
in the context of DMFT, which
puts in perspective the advantages and limitations of the new method.

\section{Rotorization}
\label{sec:rotor}

The present article emphasizes the role played by the total electron charge, and
its conjugate phase variable. We introduce a representation of the physical
electron in terms of two auxiliary fields: a fermion field which carries spin (and orbital)
degrees of freedom, and the (total) charge raising and lowering operators which
we represent in terms of a phase degree of freedom. The latter plays a role
similar to a slave boson: here a ``slave'' $O(2)$ quantum  rotor is used rather than a
conventional bosonic field.


\subsection{Atomic model}
\label{sec:atomic}

We first explain this construction on the simple example of an atomic problem,
consisting in an $N$-fold degenerate atomic level subject to a local $SU(N)$-symmetric
Coulomb repulsion:
\begin{equation}
\label{eq:Hlocal}
H_{\mr{local}} = \sum_\s \eps\, \dsd \ds + \frac{U}{2} \left[ \sum_\s \dsd \ds
  -\frac{N}{2} \right]^2
\end{equation}
We use here $\eps \equiv \epsilon_d + U/2 $ as a convenient redefinition of
the impurity level, and we recast the spin and orbital degrees of freedom into
a single index $\s = 1 \ldots N $ (for $N=2$, we have a single orbital with two
possible spin states $\s=\uparrow,\downarrow$).
We note that $\eps$ is zero at half filling, due to particle-hole symmetry.

The crucial point is that the spectrum of the atomic hamiltonian
(\ref{eq:Hlocal}) depends only on the total fermionic charge $Q=0,\cdots,N$ and has
a simple quadratic dependence on $Q$:
\begin{equation}
\label{eq:atomic_spectrum}
E_Q = \eps Q + \frac{U}{2} \left[ Q -\frac{N}{2} \right]^2
\end{equation}
There are $2^N$ states, but only $N+1$ different energy levels, with degeneracies
$\binom{N}{Q}$.
In conventional slave boson methods \cite{kotliar_ruckenstein,mott_largeN},
a bosonic field is introduced for each atomic state $|\s_1\cdots \s_Q\rangle$ (along with
spin-carrying auxiliary fermions $\fsd$). Hence, these
methods are not describing the atomic spectrum in a very economical manner.

The spectrum of (\ref{eq:Hlocal}) can actually be reproduced by introducing,
besides the set of auxiliary fermions $\fsd$,
a single additional variable, namely the angular momentum
$\hat{L} = -i \partial/\partial\theta$ associated with a quantum $O(2)$ rotor
$\theta$ ({\it i.e.} an angular variable in $[0,2\pi]$.
Indeed, the energy levels (\ref{eq:atomic_spectrum}) can
be obtained using the following hamiltonian
\begin{equation}
\label{eq:Hloc_f_theta}
H_{\mr{local}} = \sum_\s \eps \fsd \fs + \frac{U}{2} \hat{L}^2
\end{equation}
A constraint must be imposed, which insures that the total number of fermions
is equal to the $O(2)$ angular momentum (up to a shift):
\begin{equation}
\label{eq:contrainte}
\hat{L} = \sum_\s \left[\fsd \fs - \frac{1}{2} \right]
\end{equation}
This restricts the allowed values of the angular momentum to be
$l=Q-N/2=-N/2,-N/2+1,\cdots,N/2-1,N/2$, while in the absence of any
constraint $l$ can be an arbitrary (positive or negative) integer.
The spectrum of (\ref{eq:Hloc_f_theta}) is $\eps Q + Ul^2/2$, with $l=Q-N/2$
thanks to (\ref{eq:contrainte}), so that it coincides with (\ref{eq:atomic_spectrum}).

To be complete, we must show that each state in the Hilbert space can be constructed
in terms of these auxiliary degrees of freedom, in a way compatible with the Pauli
principle. This is achieved by the following identification:
\begin{equation}
\ket{\s_1\ldots \s_Q}_d \,=\,
\ket{\s_1\ldots \s_Q}_f \ket{\ell=Q-N/2}_\theta
\end{equation}
in which $\ket{\s_1\ldots \s_Q}_{d,f}$ denotes the antisymmetric fermion state
built out of $d-$ and $f-$ fermions, respectively, and $\ket{\ell}_\theta$
denotes the quantum rotor eigenstate with angular momentum $l$,
{\it i.e.} $\langle\theta \ket{\ell}_\theta = e^{i \ell \theta}$.
The creation of a physical electron with spin $\sigma$ corresponds to
acting on such a state with $\fsd$ as well as raising the total charge
(angular momentum) by one unit. Since the raising operator
is $e^{i\theta}$, this leads to the representation:
\begin{equation}
\label{eq:repres}
\dsd \equiv \fsd\, e^{i \theta}\,\,\,,\,\,\,
\ds \equiv \fs\, e^{-i\theta}
\end{equation}
Let us illustrate this for $N=2$ by writing the four possible
states in the form: $\ket{\uparrow}_d = \ket{\uparrow}_f \ket{0}_\theta$,
$\ket{\downarrow}_d = \ket{\downarrow}_f \ket{0}_\theta$,
$\ket{\uparrow\downarrow}_d = \ket{\uparrow\downarrow}_f \ket{+1}_\theta$
and $\ket{0}_d = \ket{0}_f \ket{-1}_\theta$, and showing that this
structure is preserved by $\dsd = \fsd e^{i\theta} $. Indeed:
\begin{equation}
\ket{\uparrow\downarrow}_d = d^\dagger_{\uparrow} \ket{\downarrow}_d =
f^\dagger_{\uparrow} \ket{\downarrow}_f e^{+i\theta} \ket{0}_\theta  =
\ket{\uparrow\downarrow}_f \ket{+1}_\theta
\end{equation}
The key advantage of the quantum rotor representation is that the original
quartic interaction between fermions has been replaced in (\ref{eq:Hloc_f_theta})
by a simple kinetic term ($\hat{L}^2$) for the phase field.

\subsection{Rotor representation of Anderson impurity models}
\label{sec:rotor_AIM}

We now turn to an $SU(N)$-symmetric Anderson impurity model in which
the atomic orbital is coupled to a conduction electron bath :
\begin{eqnarray}\label{eq:Anderson}
H & = &\sum_{\s} \eps \dsd \ds +
\frac{U}{2} \left[ \sum_\s \dsd \ds -\frac{N}{2} \right]^2\\
\nonumber
&  &+ \sum_{k,\s} \epsilon_k c^\dagger_{k,\s} c_{k,\s}  + \sum_{k,\s} V_k \left( c^\dagger_{k,\s} \ds
+ \dsd c_{k,\s} \right)
\end{eqnarray}
Using the representation (\ref{eq:repres}), we can rewrite this hamiltonian
in terms of the $(\fsd, \theta)$
fields only:
\begin{eqnarray}
H & = &\sum_{\s} \eps \fsd \fs +
\frac{U}{2} \hat{L}^2
+ \sum_{k,\s} \epsilon_k c^\dagger_{k,\s} c_{k,\s}  \\
\nonumber
&  & + \sum_{k,\s} V_k \left( c^\dagger_{k,\s} \fs e^{-i\theta}
+ \fsd c_{k,\s} e^{i\theta} \right)
\label{eq:Hinit}
\end{eqnarray}
We then set up a functional integral formalism for the $\fsd$ and $\theta$
degrees of freedom, and derive the action associated with (\ref{eq:Hinit}).
This is simply done by switching
from phase and angular-momentum {\it operators} $(\theta,\hat{L})$ to {\it fields}
$(\theta,\partial_\tau \theta)$ depending on imaginary time $\tau\in[0,\beta]$.
The action is constructed from $S \equiv \int_0^\beta \!\! d\tau [ - i L\,
\dt \theta + H + f^\dagger \dt f ]$, and an integration over $\hat{L}$ is performed.
It is also necessary to introduce a
complex Lagrange multiplier $h$ in order to implement the constraint $\hat{L} =
\sum_\s \fsd \fs - N/2$. We note that, because of the charge conservation on the
local impurity, $h$ can be chosen to be independent of time, with
$i h \in [0,2\pi/\beta]$.

This leads to the following expression of the action:
\begin{eqnarray}
\nonumber
S & = &\inte \sum_{\s} \fsd (\dt+\eps-h) \fs +
\frac{(\dt \theta + ih)^2}{2U} + \frac{N}{2} h \\
\label{eq:actionmoteur}
&\!\!+&\!\!\!\!\sum_{k,\s} \left[ c^\dagger_{k,\s} ( \dt + \epsilon_k) c_{k,\s}
+ V_k  c^\dagger_{k,\s} \fs e^{-i\theta}  + h.c. \right]
\end{eqnarray}
We can recast this formula in a more compact form by introducing the hybridization
function:
\begin{equation}
\Delta(i\omega) \equiv \sum_k \frac{|V_k|^2}{i\omega - \epsilon_k}
\end{equation}
and integrating out the conduction electron bath. This leads to the final form of the
action of the $SU(N)$ Anderson impurity model in terms of the auxiliary fermions and
phase field:
\begin{eqnarray}
\nonumber
S & = &\inte \sum_{\s} \fsd (\dt+\eps-h) \fs + \frac{(\dt \theta + ih)^2}{2U} + \frac{N}{2} h \\
\label{eq:action2}
& \!\!+& \!\!\!\!\inte\!\!\inte' \Delta(\tau\!-\!\tau') \sum_{\s} \fsd(\tau)\fs(\tau')\,
e^{i\theta(\tau) - i\theta(\tau')}
\end{eqnarray}

\subsection{Slave rotors, Hubbard-Stratonovich and gauge transformations}
\label{sec:gauge}
In this section, we present an alternative derivation of the expression
(\ref{eq:action2}) of the action which does not rely on the concept of slave
particles.  This has the merit to give a more explicit interpretation of the
phase variable introduced above, by relating it to a Hubbard-Stratonovich decoupling
field. This section is however not essential to the rest of the paper, and can
be skipped upon first reading.

Let us start with the imaginary time action of the Anderson impurity model
in terms of the physical electron field for the impurity orbital:
\begin{eqnarray}
\nonumber
S & = &\inte \sum_{\s} \dsd (\dt+\eps) \ds + \frac{U}{2}
\left[ \sum_\s \dsd \ds -\frac{N}{2} \right]^2 \\
& \!\!+& \!\!\!\!\inte\!\!\inte' \Delta(\tau\!-\!\tau') \sum_{\s} \dsd(\tau) \ds(\tau')
\end{eqnarray}
Because we have chosen a $SU(N)$-symmetric form for the Coulomb interaction, we can
decouple it with only one bosonic Hubbard-Stratonovitch field $\phi(\tau)$:
\begin{eqnarray}
\nonumber
S & = &\inte \sum_{\s} \dsd (\dt+\eps + i \phi(\tau) ) \ds +
\frac{\phi^2(\tau)}{2U} - i \frac{N}{2} \phi(\tau) \\
& \!\!+& \!\!\!\!\inte\!\!\inte' \Delta(\tau\!-\!\tau') \sum_{\s} \dsd(\tau) \ds(\tau')
\end{eqnarray}
Hence, a linear coupling of the field $\phi(\tau)$
to the fermions has been introduced. The idea is now to eliminate this linear coupling
for all the Fourier modes of the $\phi$-field, except that corresponding to
zero-frequency:
$\phi_0 \equiv\int_0^\beta \phi \; [2\pi]$. This can be achieved by performing
the following gauge transformation:
\begin{equation}
\dsd(\tau) = \fsd(\tau)\, e^{i \int_0^\tau \phi }\, e^{-i\phi_0 \tau/\beta}
\end{equation}
The reason for the second phase factor in this expression is that it guarantees
that the new fermion field $\fsd$ also obeys antiperiodic
boundary conditions in the path integral. It is easy to check that this change of
variables does not provide any Jacobian, so that the action simply reads:
\begin{eqnarray}
\nonumber
S & = &\inte \sum_{\s} \fsd (\dt+\eps + i\frac{\phi_0}{\beta} ) \fs +
\frac{\phi^2(\tau)}{2U} - i \frac{N}{2} \frac{\phi_0}{\beta} \\
& \!\!+& \!\!\!\!\inte\!\!\inte' \Delta(\tau\!-\!\tau') \sum_{\s} \fsd(\tau) \fs(\tau')
e^{i \int_{\tau'}^\tau [\phi -\frac{\phi_0}{\beta}] }
\end{eqnarray}
We now set:
\begin{equation}\label{eq:int_theta}
\phi(\tau) = \frac{\partial\theta}{\partial\tau} + \frac{1}{\beta}\,\phi_0
\,\,\,\,\, \left(\mbox{with:}\,\,
\phi_0 \equiv\int_0^\beta \phi \; [2\pi]\right)
\end{equation}
and notice that the
field $\theta(\tau)$ has the boundary condition $\theta(\beta) = \theta(0)
\;[2\pi]$. It therefore corresponds to an $O(2)$ quantum rotor, and the expression of the
action finally reads:
\begin{eqnarray}
\nonumber
S & \!=\! &\!\!\inte \sum_{\s} \fsd (\dt+\eps+i\frac{\phi_0}{\beta}) \fs
+ \frac{(\dt \theta + \frac{\phi_0}{\beta})^2}{2U}
- i  \frac{N}{2} \frac{\phi_0}{\beta} \\
& \!\!+& \!\!\!\!\inte\!\!\inte' \Delta(\tau\!-\!\tau') \sum_{\s} \fsd(\tau) \fs(\tau')
e^{i\theta(\tau) - i\theta(\tau')}
\end{eqnarray}
This is exactly expression (\ref{eq:action2}), with the identification:
$\phi_0/\beta \equiv i h$. This, together with (\ref{eq:int_theta}), provides an
explicit relation between the quantum rotor and Lagrange multiplier fields on one side,
and the Hubbard-Stratonovich field conjugate to the total charge, on the other.

\section{Sigma-model representation and solution in the limit of many components}
\label{sec:NCA}

\subsection{From quantum rotors to a sigma model}
\label{sec:sigma_model}

Instead of using a phase field to represent the $O(2)$ degree of freedom, one can
use a constrained (complex) bosonic field $X \equiv e^{i\theta}$ with:
\begin{equation}
\left| X(\tau) \right|^2 = 1
\end{equation}
The action (\ref{eq:action2}) can be rewritten in terms of this field, provided
a Lagrange multiplier field $\lambda(\tau)$ is used to implement this constraint:
\begin{eqnarray}
\nonumber
S & = &\inte \sum_{\s} \fsd (\dt+\eps-h) \fs  + \frac{N}{2} h - \frac{h^2}{2U}  \\
\nonumber
& +  &  \inte \frac{|\dt X|^2}{2U} +\frac{h}{2U} (X^*\dt X \!\! - h.c.)
+ \lambda(\tau) (|X|^2 \!-\! 1) \\
\label{eq:action_de_phase}
& \!\!+& \!\!\!\!\inte\!\!\inte' \Delta(\tau\!-\!\tau') \sum_{\s} \fsd(\tau) \fs(\tau')
X(\tau)X^*(\tau')
\end{eqnarray}
Hence, the Anderson model has been written
as a theory of auxiliary fermions coupled to a non-linear
$O(2)$ sigma-model, with a constraint (implemented by $h$) relating
the fermions and the sigma-model field $X(\tau)$.

A widely-used limit in which sigma models become solvable, is the limit of a
large number of components of the field. This motivates us to generalize
(\ref{eq:action_de_phase}) to a model with an $O(2M)$ symmetry.
The bosonic field $X$ is thus extended to an $M$-component complex field
$X_\alpha$ ($\alpha=1\ldots M$) with $\sum_\alpha |X_\alpha|^2=M$.
The corresponding action reads:
\begin{eqnarray}
\nonumber
S & = &\inte \sum_{\s} \fsd (\dt+\eps-h) \fs  + \frac{N}{2} h - M \frac{h^2}{2U} -M\lambda \\
\nonumber
&\!\! + \!\! & \!\!\! \inte \sum_{\alpha} \frac{|\dt X_\alpha|^2}{2U} +\frac{h}{2U}
(X^*_\alpha \dt X_\alpha \!-\! h.c.) + \lambda |X_\alpha|^2  \\
\nonumber
\label{eq:action_de_phase_OM}
& \!\!+& \!\!\!\!\inte\!\!\inte' \frac{1}{M}\Delta(\tau\!-\!\tau')
\sum_{\s,\alpha} \fsd(\tau) \fs(\tau')
X_\alpha(\tau)X^*_\alpha(\tau')
\end{eqnarray}
Let us note that this action corresponds to the hamiltonian:
\begin{eqnarray}
\nonumber
H & = & \sum_{\s} \eps \fsd \fs +
\frac{U}{2M} \sum_{\alpha,\beta} \hat{L}^2_{\alpha,\beta}
+ \sum_{k,\s\alpha} \epsilon_k c^\dagger_{k,\s\alpha} c_{k,\s\alpha}  \\
\label{eq:Anderson_OM}
&  & + \sum_{k,\s,\alpha} \frac{V_k}{\sqrt{M}} \left( c^\dagger_{k,\s\alpha} \fs X^*_\alpha
+ \fsd c_{k,\s\alpha} X_\alpha \right)
\end{eqnarray}

In this expression, $\hat{L}_{\alpha,\beta}$ denotes the angular momentum tensor
associated with the $X_\alpha$ vector.
The hamiltonian (\ref{eq:Anderson_OM}) is a generalization of the $SU(N)\times O(2)=U(N)$
Anderson impurity model to an $SU(N)\times O(2M)$ model in which the total electronic charge
is associated with a specific component of $\hat{L}$. It reduces to the usual Anderson model
for $M=1$.

In the following, we consider the limit where both $N$ and $M$ become large,
while keeping a fixed ratio $N/M$. We shall demonstrate that exact coupled integral
equations can be derived in this limit, which determine the Green's functions of the
fermionic and sigma-model fields (and the physical electron Green's function as well).
The fact that these coupled integral equations do correspond to the exact solution of a
well-defined hamiltonian model (Eq.~(\ref{eq:Anderson_OM})) guarantees that no unphysical
features (like e.g violation of causality) arise in the solution.
Naturally, the generalized hamiltonian (\ref{eq:Anderson_OM}) is a formal extension of the
Anderson impurity model of physical interest.
Extending the charge symmetry from $O(2)$ to $O(2M)$ is not entirely inocuous, even at
the atomic level: as we shall see below, the energy levels of a single $O(2M)$ quantum
rotor have multiple degeneracies, and depend on the charge (angular momentum) quantum
number in a way which does not faithfully mimic the $O(2)$ case. Nevertheless,
the basic features defining the generalized model (a localized orbital subject to a
Coulomb charging energy and coupled to
an electron bath by hybridization) are similar to the original model of physical interest.

\subsection{Integral equations}
\label{sec:integral_equations}

In this section, we derive coupled integral equations which become
exact in the limit where {\it both} $M$ and $N$
are large with a fixed ratio:
\begin{equation}\label{eq:limit}
\cN \equiv \frac{N}{M}\,\,\,\,\,(N,M\rightarrow\infty)
\end{equation}
Following \cite{OP_AG_GK_AS}
(see also \cite{Cox_Ruckenstein}),
the two body interaction
between the auxiliary fermions and the sigma-model fields is decoupled using
(bosonic) bi-local fields $Q(\tau,\tau')$ and
$\qbar(\tau,\tau')$ depending on two times. Hence, we consider the action:
\begin{eqnarray}
\nonumber
S & = &\inte \sum_{\s} \fsd (\dt+\eps-h) \fs  + \frac{N}{2} h - M \frac{h^2}{2U} -M\lambda \\
\nonumber
&\!\! + \!\! & \!\!\! \inte \sum_{\alpha} \frac{|\dt X_\alpha|^2}{2U} +\frac{h}{2U}
(X^*_\alpha \dt X_\alpha \!-\! h.c.) + \lambda |X_\alpha|^2  \\
\nonumber
& + &  \inte \inte' \; M \frac{Q(\tau,\tau') \qbar(\tau,\tau')}{\Delta(\tau-\tau')}\\
\nonumber
& - &  \; \inte \inte' \; \qbar(\tau,\tau') \sum_\alpha X_\alpha(\tau) X_\alpha^*(\tau') \\
\label{eq:actionlargeNM}
& + &  \inte \inte' \; Q(\tau,\tau') \sum_{\s} \fsd(\tau) \fs(\tau')
\end{eqnarray}
In the limit (\ref{eq:limit}), this action is controlled by a saddle-point,
at which the Lagrange multipliers take static expectation values $h$ and $\lambda$, while
the saddle point values of the $Q$ and $\qbar$ fields are translation invariant functions
of time $Q(\tau-\tau')$ and $\qbar(\tau-\tau')$.

Introducing the imaginary-time Green's functions of the auxiliary fermion
and sigma-model fields as:
\begin{eqnarray}
\label{eq:Gfdef}
G_f(\tau) & \equiv & - \left<T_\tau \fs(\tau) \fsd(0) \right> \\
\label{eq:GXdef}
G_X(\tau) & \equiv & + \left<T_\tau X_\alpha(\tau) X^*_\alpha(0) \right>
\end{eqnarray}
we differentiate the effective action (\ref{eq:actionlargeNM}) with
respect to $Q(\tau)$ and $\qbar(\tau)$, which leads to the following saddle-point equations:
$\qbar(\tau) = - \cN \Delta(\tau) G_f(-\tau)$ and $Q(\tau) = \Delta(\tau) G_X(\tau)$.
The functions $Q(i\omega_n)$ ($=\Sigma_f$) and $\qbar^*(i\nu_n)$ ($=\Sigma_X$) define
fermionic and bosonic self-energies:
\begin{eqnarray}
\label{eq:Gf}
G^{-1}_f(i\omega_n) & = & i\omega_n -\eps+h - \Sigma_f(i\omega_n)\\
\label{eq:GX}
G^{-1}_X(i\nu_n) & = & \frac{\nu_n^2}{U}+\lambda -\frac{2 i h\nu_n}{U} -\Sigma_X(i\nu_n)
\end{eqnarray}
where $\omega_n$ ({\it resp.} $\nu_n$) is a fermionic ({\it resp.} bosonic)
Matsubara frequency.
The saddle point equations read:
\begin{eqnarray}
\label{eq:col1}
\Sigma_X(\tau) & = & - \cN \Delta(-\tau) G_f(\tau) \\
\label{eq:col2}
\Sigma_f(\tau) & = & \Delta(\tau) G_X(\tau)
\end{eqnarray}
together with the constraints associated with $h$ and $\lambda$:
\begin{eqnarray}
\label{eq:col3}
G_X(\tau\myeq0) & = & 1 \\
\label{eq:col4}
G_f(\tau\myeq0^-) & = & \frac{1}{2} - \frac{2h}{\cN U} \\
\nonumber
& + & \frac{1}{\cN U} \left[ \partial_\tau G_X(\tau\myeq0^-) + \partial_\tau
G_X(\tau\myeq0^+) \right]
\end{eqnarray}
There is a clear similarity between the structure of these coupled integral
equations and the infinite-$U$ NCA equations \cite{NCA}. We note also significant
differences, such as the constraint equations. Furthermore, the finite value of
the Coulomb repulsion $U$ enters the bosonic propagator (\ref{eq:GX}) in a quite novel manner.

The two key ingredients on which the present method are based is the use of a slave rotor
representation of fermion operators, and the use of integral equations for the frequency-dependent
self-energies and Green's functions. For this reason, we shall denote the integral equations
above under the name of ``Dynamical Slave Rotor'' method (DSR) in the following.

\subsection{Some remarks}
\label{sec:remarks}

We make here some technical remarks concerning these integral equations.

First, we clarify how the interaction parameter $U$ was scaled in order to
obtain the DSR equations above. This issue is related to the manner in which
the atomic limit ($\Delta=0$) is treated in this method. In the original $O(2)$
atomic hamiltonian (\ref{eq:Hlocal}), the charge gap between the ground-state and
the first excited state is $U/2$ at half-filling ($\epsilon_0=0$).
In the DSR method, the charge gap is associated with the gap in the
slave rotor spectrum. If the $O(2M)$ generalization of (\ref{eq:Hlocal})
is written as in (\ref{eq:Anderson_OM}):
\begin{equation}
H_{int} = \frac{U}{2M} \sum_{\alpha,\beta} \hat{L}^2_{\alpha,\beta}
\label{eq:rotor_OM}
\end{equation}
the spectrum reads:
$E_\ell = U \ell (\ell + 2M-2)/(2M)$. As a result, the energy difference from the
ground state to the first excited state
is $ E_1 - E_0 =  U \, (2M-1)/(2M) \simeq U$ at large $M$, whereas it is $U/2$ at $M=1$.
In order to use the DSR method in practice as an approximate impurity solver,
the parameter $U$ should thus be normalized in a different way than in
(\ref{eq:Anderson_OM}), so that the gap is kept equal to $U/2$ in the
large-M limit as well.
Technically this can be enforced
by choosing the following normalization:
\begin{equation}
\label{eq:normalization}
H_{int}' = \frac{U}{4M-2} \sum_{\alpha,\beta} \hat{L}^2_{\alpha,\beta}
\end{equation}
instead of (\ref{eq:rotor_OM}).
Note that this scaling coincides with (\ref{eq:rotor_OM}) for $M=1$, but does yield
$E_1 - E_0=U/2$ for large-M, as desired.
This definition of $U$ was actually used when writing the saddle-point integral equations
(\ref{eq:GX}), although we postponed the discussion of this point to the present section
for reasons of simplicity.

Let us elaborate further on the accuracy of the DSR integral equations in the
atomic limit. In Appendix~\ref{app:atomic}, we show that the physical
electron spectral function obtained within DSR in the atomic limit coincides
with the exact $O(2)$ result at half-filling and at $T=0$. This is a non-trivial
result, given the fact that the constraint is treated on average and the above
remark on the spectrum spectrum.
In contrast to NCA, the DSR method (in its present form) is not based by construction
on a strong-coupling expansion around the exact atomic spectrum, so that this is
a crucial check for the applicability of this method in practice.
In the context of DMFT for example, it is essential in order to describe correctly the Mott
insulating state \cite{IPT_Mott}.
However, the DSR integral equations fail to reproduce exactly the O(2) atomic limit
off half-filling, as explained in Appendix~\ref{app:atomic}. Deviations become severe
for too high dopings, as discussed in Sec.~\ref{sec:doping}.
This makes the present form
of DSR applicable only for systems in the vicinity of half-filling.

We now discuss some general spectral properties of the DSR solver.
From the representation of the physical electron field $\dsd \myeq \fsd X$,
and from the convention chosen for the pseudo-particules
Green's functions (\ref{eq:Gfdef}-\ref{eq:GXdef}), the one-electron
physical Green's function is simply expressed as:
\begin{equation}
\label{eq:green}
G_d(\tau) = G_f(\tau) G_X(-\tau)
\end{equation}
Therefore eq. (\ref{eq:col3}), combined with the fact that $\fsd$ has a $(-1)$
discontinuity at $\tau=0$ (which is obvious from (\ref{eq:Gf})), shows that $\dsd$
possesses also a $(-1)$ jump at zero imaginary time. This ensures that the physical
spectral weight is unity in our theory, and thus that physical spectral
function are correctly normalized. Because the DSR integral equations
result from a controlled large $N,M$ limit,
it also insures that the physical self-energy always
have the correct sign ({\it i.e.} $\mcal{I}m \, \Sigma_d(\omega+i0^+) < 0 $).
This is {\it not} the case \cite{pathologies_U_NCA} for the
finite $U$ version of the NCA \cite{U_NCA} which is constructed as a
resummation of the strong-coupling expansion in the hybridization function
$\Delta(\tau)$ (Strong coupling
resummations generically suffer from non-causality, see also
\cite{hubbard_strong_coupling} for an illustration.)

Finally, we comment on the non-interacting limit $U\rightarrow0$. This is a major failure
of the usual NCA, which limits its applicability in the weakly correlated
regime.  In the DSR formalism, this limit is {\it exact} as can be noticed
from equation (\ref{eq:GX}).
Indeed, as $U$ vanishes, only the zero-frequency component of $G_X(i\nu_n)$ survives,
so that $G_X(\tau)$ simply becomes a constant. Because of the constraint
(\ref{eq:col3}), we get correctly $G_X(\tau)=1$ at $U=0$. From (\ref{eq:col2}) and
(\ref{eq:green}), this proves that $G_d(\tau)$ is the non-interacting Green's function:
\begin{equation}
G^{U=0}_d(i\w_n) = \frac{1}{i\w_n - \epsilon_0 - \Delta(i\w_n)}
\end{equation}

We finally acknowledge that an alternative dynamical approximation to the finite $U$ Anderson
model \cite{NCA_FLT} was recently developed as an extension of NCA by Kroha, W\"{o}lfle and
collaborators (a conventional slave boson representation was used in this work).
Many progresses have been made following this method, but, to the authors' knowledge, this
technique has not yet been implemented in the context of DMFT (one of the
reasons is its computational cost).
By developing the DSR approximation, we pursue a rather complementary goal: the aim here is
not to improve the low-energy singularities usually encountered with integral equations,
but rather to have a fast and efficient solver which reproduces correctly the main features
of the spectral functions and interpolates between weak and strong coupling. In that sense,
it is very well adapted to the DMFT context.

\section{Application to the single-impurity Anderson model}
\label{sec:SIAM}

We now discuss the application of the DSR in the simplest setting: that of a single
impurity hybridized with a fixed bath of conduction electrons.
For simplicity, we focus on the half-filled, particle-hole
symmetric case, which implies $\eps = h = 0$.
The doped (or mixed valence) case will be addressed in the next section, in the
context of DMFT.

As the strength of the Coulomb interaction $U$ is increased from weak to strong
coupling, two well-known effects are expected (see e.g \cite{hewson}).
First the width of the low-energy resonance is reduced from its non-interacting
value $\Delta_0 \equiv |\Delta''(0)|$. As one enters the Kondo regime
$\Delta_0 \ll U \ll\Lambda $ (with $\Lambda$ the conduction electrons bandwidth),
this width becomes a very small energy scale, of the order of the Kondo
temperature:
\begin{equation}
T_K \myeq \sqrt{2 U \Delta_0} \exp(-\pi U/ (8 \Delta_0))
\label{eq:tk}
\end{equation}
A (local) Fermi liquid description applies, with quasiparticles having a
large effective mass and small weight: $Z=m/m^* \sim T_K/\Delta_0$. The
impurity spin is screened for $T<T_K$.

Second, the corresponding spectral weight is transferred to high energies,
into ``Hubbard bands'' associated to the atomic-like transitions
(adding or removing an electron into the half-filled impurity orbital),
broadened by the hybridization to the conduction electron bath.
The suppression of the low-energy spectral weight corresponds to the
suppression of the charge fluctuations on the local orbital.
These satellites are already visible at moderate values of the coupling
$U/\Delta_0$. As temperature is increased from $T<T_K$ to $T>T_K$, the
Kondo quasiparticle resonance is quickly destroyed, and the missing
spectral weight is added to the Hubbard bands.

The aim of this section is to investigate whether the integral equations
introduced in this paper reproduce these physical effects in a satisfactory
manner.

\subsection{Spectral functions}

We have solved numerically these integral equations by iteration, both on the imaginary
axis and for real frequencies. Working on the imaginary axis is technically much
easier. A discretization of the interval $\tau \in [0,\beta]$ is used (with typically 8192
points, and up to 32768 for reaching the lowest temperatures), as well as Fast 
Fourier Transforms for the Green's functions.
Searching by dichotomy for the saddle-point value of the Lagrange multiplier $\lambda$
(Eq.~\ref{eq:col3}) is conveniently implemented at each step of
the iterative procedure.
Technical details about the analytic continuation of (\ref{eq:col1}-\ref{eq:col4})
to real frequencies, as well as their numerical solution are
given in Appendix~\ref{appendix1}.

\begin{figure}[htbp]
\begin{center}
\vspace{0.6cm}
\includegraphics[width=7cm]{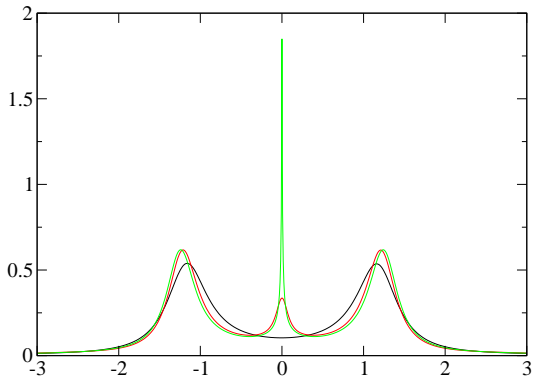}
\end{center}
\caption{$d$-level spectral function $\rho_d(\w)$ for fixed $U = 2$
and inverse temperatures $\beta = 4,20,2000$, with the conduction electron bath
described in the text}
\label{fig:pousse_Kondo}
\end{figure}

\begin{figure}[htbp]
\begin{center}
\vspace{0.6cm}
\includegraphics[width=7cm]{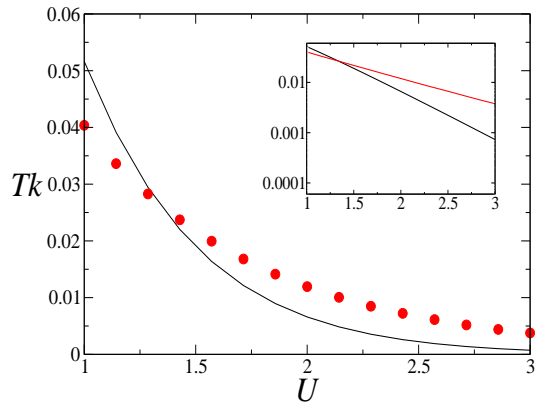}
\end{center}
\caption{Kondo temperature $T_K$ from the exact formula (line) and from the numerical
solution of the DSR equations (dots). Units of energy are such that $\Delta_0=0.16$.}
\label{fig:compare_TK}
\end{figure}

On Fig.~\ref{fig:pousse_Kondo}, we display our results for the 
impurity-orbital spectral function $\rho_d(\omega)$, at three different
temperatures (the density of states of the conduction electron bath
is chosen as a semi-circle with half-width $\Lambda = 6$, the
resonant level width is $\Delta_0= 0.16$ and we take $U = 2$).
The growth of the Kondo resonance as the temperature is lowered is
clearly seen. The temperatures in Fig.~\ref{fig:pousse_Kondo} have been
chosen such as to illustrate three different regimes: for $T\gg T_K$, no
resonance is seen and the spectral density displays a ``pseudogap''
separating the two high-energy bands; for $T\simeq T_K$ transfer of spectral weight
to low energy is seen, resulting in a fully developed Kondo resonance
for $T\ll T_K$.
We have not obtained an analytical determination of the Kondo temperature
within the present scheme. In the case of NCA equations, it is possible to
derive a set of differential equations in the limit of infinite bandwith which
greatly facilitate this. This procedure cannot be applied here,
because of the form (\ref{eq:GX}) of the boson propagator. Nevertheless, we checked
that the numerical estimates of the width of the Kondo peak is indeed
exponentially small in $U$ as in formula (\ref{eq:tk}), see Fig.~\ref{fig:compare_TK}. 
However, because $U$ is normalized as in (\ref{eq:normalization}), (which gives the 
correct atomic limit), the prefactor inside the exponential appears to be twice too small.

\begin{figure}[htbp]
\begin{center}
\vspace{0.6cm}
\includegraphics[width=7cm]{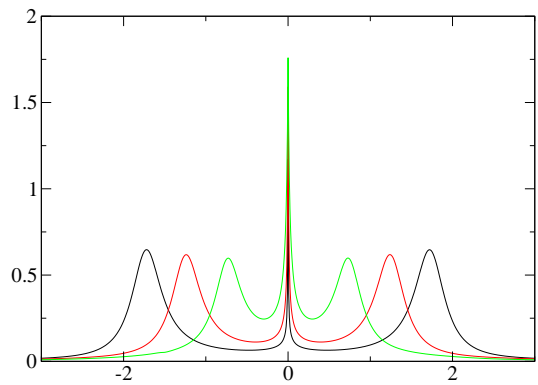}
\end{center}
\caption{$\rho_d(\omega)$ at low temperature ($\beta=600$) and for increasing
$U=1,2,3$}
\label{fig:analyse_TK}
\end{figure}

On Fig.~\ref{fig:analyse_TK}, we display the spectral function for a
fixed low temperature and increasing values of $U$.
The strong reduction of the Kondo scale (resonance width) upon increasing $U$
is clear on this figure.
We note that the high-energy peaks have a width which remains of order
$\Delta_0$, independently of $U$, which is satisfactory. However, we also note
that they are not peaked exactly at the atomic value $\pm U/2$, which might
be an artefact of these integral equations. The shift is rather small however.

\begin{figure}[htbp]
\begin{center}
\vspace{0.6cm}
\includegraphics[width=7cm]{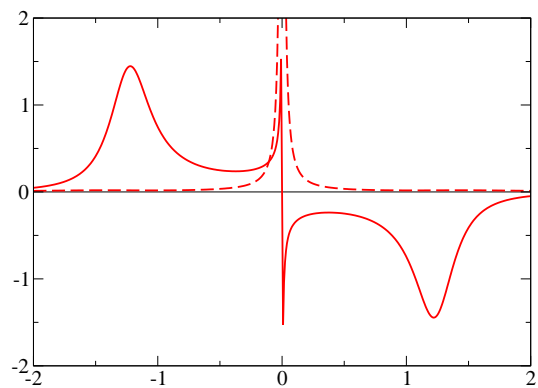}
\end{center}
\caption{Pseudo-particule spectral function at $U=2$ and
$\beta=600$. The Kondo resonance is visible in $\rho_f(\w)$, broken curve, whereas
$\rho_X(\w)$ displays higher energy features, plain curve}
\label{fig:pseudo}
\end{figure}

Fig.~\ref{fig:pseudo} illustrates how the high- and low- energy features in the
d-level spectral functions are associated with corresponding features in
the auxiliary particle spectral functions $\rho_f$ and $\rho_X$. In particular,
the sigma-model boson (slave rotor) is entirely responsible for the Hubbard
bands at high energy (as expected, since it describes charge fluctuations).

\begin{figure}[htbp]
\begin{center}
\vspace{0.6cm}
\includegraphics[width=7cm]{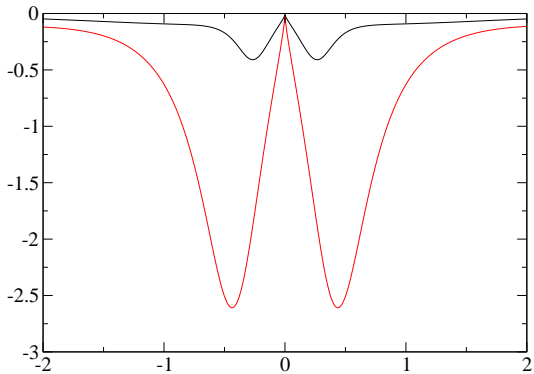}
\end{center}
\caption{Imaginary part of the physical self-energy on the real-frequency axis, for
$U=0.5$ (upper curve) and $U=2$ (lower curve) at $\beta=600$}
\label{fig:sd}
\end{figure}

As stressed in the introduction, an advantage of our scheme is that the d-level
self-energy is always causal, even for small $U$ or large doping. This is
definitely an improvement as compared to the usual $U$-NCA approximation.
This is illustrated by Fig.~\ref{fig:sd}, from which it is also clear
that $\Sigma_d$ decreases (and eventually vanishes) as $U$ goes to zero
(for a more detailed discussion and comparison to NCA, see Sec.~\ref{sec:NCA}).
However, it is also clear from this figure that the low-energy behavior of the
self-energy is not consistent with Fermi liquid theory.
This is a generic drawback of NCA-like integral
equation approaches, that we now discuss in more details.

\subsection{Low-energy behaviour and Friedel sum-rule}
\label{sec:Friedel}

We discuss here the low-energy behavior of the integral equations (in the case of
a featureless conduction electron bath), for both the d-level and auxiliary
field Green's functions.
As explained below, this low-energy behavior depends sensitively on the ratio
$\cN=N/M$ which is kept fixed in the limit considered in this paper.
The calculations are detailed in Appendix~\ref{app:low}, where we establish the
following.

\begin{itemize}
\item {\it Friedel sum rule}. The zero-frequency value of the d-level spectral function at
$T=0$ is independent of $U$ and reads:
\begin{eqnarray}
\nonumber
\rho_d(\omega=T=0) &=& \frac{-1}{\pi \Delta''(0)}
\frac{\pi/2}{\cN+1}\tan\left(\frac{\pi}{2} \frac{\cN}{\cN+1}\right)\\
\label{eq:friedel}
\end{eqnarray}
This is to be contrasted with the exact value for the $O(2)$ model ($M=1$), which is
independent of $N$ and reads:
\begin{equation}
\rho_d^{\mr{exact}} (\omega=T=0)= \frac{-1}{\pi \Delta''(0)} \;\;\;  (M=1, \;\;
\mr{any} \; N)
\label{eq:exact}
\end{equation}
This is the {\it Friedel sum rule} \cite{yamada,hewson}, which in this particle-hole
symmetric case simply follows from the Fermi-liquid requirement that the (inverse)
lifetime $\Sigma''_d(\omega=0)$ should vanish at $T=0$ (since
$G_d^{-1}=i\omega-\Delta(i\omega)-\Sigma_d(i\omega)$). As a result,
$\rho_d(0)$ is pinned at its non-interacting value.
The integral equations discussed here yield a non-vanishing $\Sigma''_d(\omega=0)$ (albeit
always negative in order to satisfy causality), and
hence do not describe a Fermi liquid at low energy.
The result (\ref{eq:friedel}) is
identical to that found in the NCA for $U=\infty$, but
holds here for arbitrary $U$.
There is actually no contradiction between this remark and the fact
that our integral equations yield the exact spectral function in the
$U\rightarrow~0$ limit. Indeed, the limit $U\rightarrow 0$ at finite $T,\omega$ does
not commute with $\omega,T\rightarrow 0$ at finite $U$
(in which (\ref{eq:friedel}) holds).

\item{\it Low-frequency behavior.} The auxiliary particle spectral functions have a
low-frequency singularity characterized by exponents which depend continuously on $\cN$
(as in the $U=\infty$ NCA):
$\rho_f(\omega)\propto 1/\Sigma''_f(\omega)\propto 1/|\omega|^{\alpha_f}$,
$\rho_X(\omega)\propto 1/\Sigma''_X(\omega)\propto \mr{Sign}(\omega)/|\omega|^{\alpha_X}$,
with: $\alpha_f=1-\alpha_X=1/(\cN+1)$.
These behavior are characterized more precisely in Appendix~\ref{app:low}.
A power-law behavior is also found for the
physical self-energy $\Sigma''_d(\omega)-\Sigma''_d(0)$ at low frequency
(as evident from Fig.~\ref{fig:sd}).

\end{itemize}

Let us comment on the origin of these low-energy features, as well as on their
consequences for the practical use of the present method.

First, it is clear from expression (\ref{eq:Anderson_OM}) that the Anderson impurity
hamiltonian generalized to $\mr{SU}(N)\times \mr{O}(2M)$ actually involves {\it $M$ channels of
conduction electrons}. Hence, the non-Fermi liquid behaviour found when solving the
integral equations associated with the $N,M\rightarrow \infty$ limit simply follows from the
fact that multi-channel models lead to {\it overscreening} of the impurity spin, and correspond
to a non-Fermi liquid fixed point. In that sense, these integral equations reproduce very
accurately the expected low-energy physics, as previously studied
for the simplest case of the Kondo model in \cite{Cox_Ruckenstein,OP_AG_GK_AS}.

Naturally, this means that the use of such integral equations to describe the one-channel
(exactly screened) case becomes problematic in the low-energy region. In particular, the
exact Friedel sum rule is violated, the d-level lifetime remains finite at low energy and
non-Fermi liquid singularities are found. While the approach is reasonable in order
to reproduce the overall features of the one-electron spectra, it should not be employed
to calculate transport properties at low energy for example. We note however that the
deviation from the exact Friedel sum rule vanishes in the $\cN\rightarrow \infty$ limit.
This is expected from the fact that in this limit the number of channels ($M$) is small as
compared to orbital degeneracy ($N$). The violation of the sum rule remains rather small
even for reasonable values of $\cN$. This parameter can actually be used as an adjustable
parameter when using the present method as an approximate impurity solver.
There is no fundamental reason for which $\cN=N$ should provide the best
approximate description of the spectral functions of the one-channel case.
We shall use this possibility when applying this method in the DMFT context
in the next section: there, we choose $\cN=3$ in order to adjust to the known critical
value of $U$ for a single orbital. We also used this value in the calculations reported
above. The zero-frequency value of the spectral density
is thus $\rho_d(0)\simeq 1.85$, while the Friedel sum rule would
yield $\rho_d(0)\simeq 1.95$ (cf. Fig.~\ref{fig:analyse_TK}).
Hence the violation of the sum rule is a small effect (of the order of $5\%$),
comparable to the one found with "numerically exact" solvers,
due to discretization errors \cite{NRG}.
Also, we point out that the pinning of $\rho_d(0)$ at a value independent of $U$
(albeit not that of the Friedel sum-rule) is an important aspect of the present
method, which
will prove to be crucial in the context of DMFT in order to recover the correct
scenario for the Mott transition.

Finally, we emphasize that increasing the parameter $\cN$ also corresponds to
increasing the orbital degeneracy of the impurity level. This will be studied in
more details in Sec.~\ref{sec:multi-orbital}. In particular, we shall see that correlation
effects become weaker as $\cN$ is increased (for a given value of $U$), due to
enhanced orbital fluctuations.

\section{Applications to Dynamical Mean-Field Theory and the Mott transition}
\label{sec:DMFT}

\subsection{One-orbital case: Mott transition, phase diagram}
\label{sec:Mott_one_orbital}

Dynamical Mean-Field Theory has led to significant progress in our understanding of
the physics of a correlated metal close to the Mott transition \cite{RMP_DMFT}.
The detailed description of this transition itself within DMFT is now well
established \cite{RMP_DMFT,IPT_Mott,Moeller,Landau_functional,QMC_Mott,NRG,mott_largeN}.
In this section, we use these established results as a benchmark and test
the applicability of the method introduced in this paper in the context of DMFT, with
very encouraging results.
As explained above, this is particularly relevant in view of the recent applications
of DMFT to electronic structure calculations of correlated solids
\cite{LDA_DMFT}, which call for efficient multi-orbital impurity solvers.

As is well-known \cite{IPT,RMP_DMFT,pruschke-review}, DMFT maps a lattice hamiltonian onto a self-consistent
quantum impurity model. We discuss first the half-filled Hubbard model,
and address later the doped case. We then have to solve a
particle-hole symmetric Anderson impurity model :
\begin{eqnarray}
S & = &\inte \sum_{\s} \dsd \dt \ds +
\frac{U}{2} \left[ \sum_\s \dsd \ds - 1 \right]^2 \\
\nonumber
& + & \inte \inte' \; \Delta(\tau-\tau') \sum_{\s} \dsd(\tau) \ds(\tau')
\end{eqnarray}
subject to the self-consistency condition:
\begin{equation}
\label{eq:self}
\Delta(\tau) = t^2  G_d(\tau)
\end{equation}
In this expression, a semi-circular density of states with half-bandwith
$D = 2t$ has been considered, corresponding to a infinite-connectivity Bethe
lattice ($z=\infty$) with hopping $t_{ij}=t/\sqrt{z}$. In the following, we shall
generally express all energies in units of $D$ ($D=1$).

In practice, one must iterate numerically the ``DMFT loop'':
$\Delta(\tau) \rightarrow G_d(\tau) \rightarrow \Delta(\tau)_{new}=t^2G_d$,
using some ``impurity solver''.  Here, we make use of the
integral equations (\ref{eq:col1}-\ref{eq:col4}).
The hybridization function $\Delta(\tau)$ being determined by the
self-consistency condition
(\ref{eq:self}), there are only two free parameters, the local Coulomb
repulsion $U$ and the temperature $T$ (normalized by $D$).

\begin{figure}[htbp]
\begin{center}
\vspace{0.6cm}
\includegraphics[width=7cm]{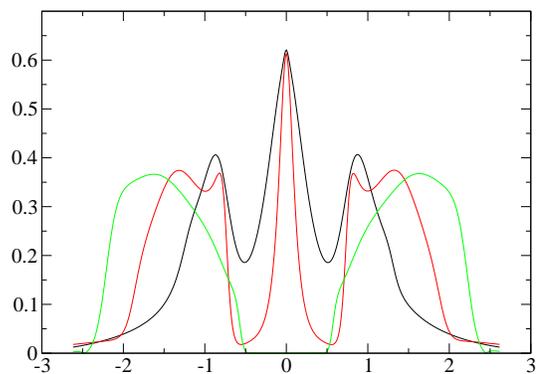}
\end{center}
\caption{Local spectral function at $\beta=40$ and $U=1, 2, 3$ for the
half-filled Hubbard model within DMFT, as obtained with the DSR solver }
\label{fig:transition}
\end{figure}

\begin{figure}[htbp]
\begin{center}
\vspace{0.6cm}
\includegraphics[width=7cm]{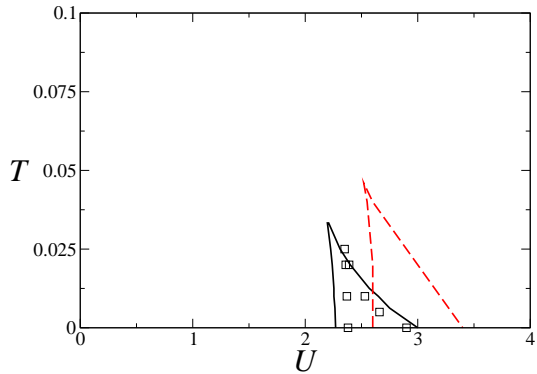}
\end{center}
\caption{Single-orbital (paramagnetic) phase diagram at half-filling. Squares indicate the ED
result, the DSR result is the solid line, and IPT the broken line}
\label{fig:phase}
\end{figure}

We display in Fig.~\ref{fig:transition} the spectral functions obtained at
low temperature, for increasing values of $U$, and in
Fig.~\ref{fig:phase} the corresponding phase diagram. The value of the
parameter $\cN$ has been adapted to the description of the one-orbital case (see below).
The most important point is that we find a {\it coexistence region} at low-enough
temperature: for a range of couplings $U_{c1}(T)\leq U \leq U_{c2}(T)$, both a
metallic solution and an insulating solution of the (paramagnetic) DMFT equations
exist. The Mott transition is thus first-order at finite temperatures.
This is in agreement with the results established for this problem
by solving the DMFT equations with controlled numerical methods
\cite{QMC_Mott,NRG}, as well as with analytical results
\cite{Landau_functional,Moeller}.
In particular, the spectral functions that we obtain (Fig.~\ref{fig:transition})
display the well-known
separation of energy scales found within DMFT: there is a
gradual narrowing of the quasiparticle peak, together with a preformed Mott gap
at the transition. In the next section, we compare these spectral functions
to those obtained using other approximate solvers. As pointed out there,
despite some formal similarity in the method, it is well-known that the
standard $U$-NCA does not reproduce correctly this separation
of energy scales close to the transition \cite{jarrell_NCA}.

Let us explain how the parameter $\cN$ has been chosen in these calculations.
As demonstrated below, the values of the critical couplings $U_{c1}$ and
$U_{c2}$ (and hence the whole phase diagram and coexistence window)
strongly depends on the value of this parameter. This is expected, since
$\cN$ is a measure of orbital degeneracy. What has been done in the calculations
displayed above is to choose $\cN$ in such a way that the known value \cite{Moeller}
$U_{c2}(T=0) \simeq 2.9$ of the critical coupling at which the $T=0$ metallic solution
disappears in the single-orbital case, is accurately reproduced.
We found that this requires $\cN \simeq 3$ (note that $N/M=2$ in the one-orbital case,
so that the best agreement is not found by a naive application of the large $N,M$
limit). This value being fixed, we find a
critical coupling $U_{c1}(T=0)\simeq 2.3$ in good agreement with the
value from (adaptative) exact diagonalizations $U_{c1}\simeq 2.4$.
The whole domain of coexistence in the
$(U,T)$ plane is also in good agreement with established results
(in particular we find the critical endpoint at $T_c\simeq 1/30$, while
QMC yields $T_c\simeq 1/40$). These are very stringent
tests of the applicability of the present method, since we have allowed ourselves to use
only one adjustable parameter ($\cN$). In Sec.\ref{sec:multi-orbital}, we study how the
Mott transition depends on the number of orbitals, which further validates the procedure
followed here.

\subsection{Comparison to other impurity solvers: spectral functions}
\label{sec:comp_solvers}

\begin{figure}[htbp]
\begin{center}
\vspace{0.6cm}
\includegraphics[width=7cm]{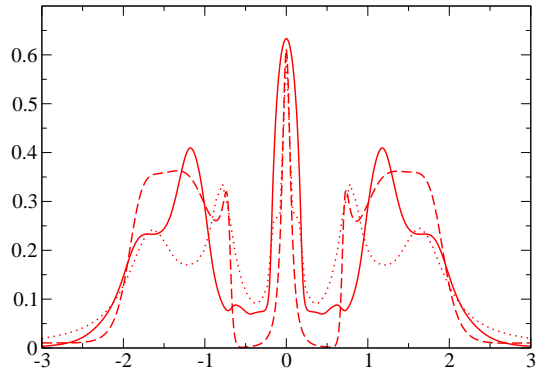}
\end{center}
\caption{Comparison between DSR (dash), IPT (straight) and ED (dot) at $U=2.4$ and $\beta=60$}
\label{fig:compare_dos_U2.4}
\end{figure}

Let us now compare the spectral functions obtained by the present method with
other impurity solvers commonly used for solving the DMFT equations.
We start with the iterated perturbation theory approximation (IPT) and
the exact diagonalization method (ED).
Both methods have played a major role in the early developments of the
DMFT approach to the Mott transition \cite{IPT_Mott}.
A comparison of the spectral functions
obtained by the present method to those obtained with IPT and ED is displayed in
Fig.~\ref{fig:compare_dos_U2.4}, for a value of $U$ corresponding to a correlated metal
close to the transition.

The overall shape and characteristic features of the spectral function are quite similar
for the three methods. A narrow quasiparticle peak is formed, together with Hubbard bands,
and there is a clear separation of energy scales between the width of the central peak
(related to the quasiparticle weight) and the ``preformed'' Mott (pseudo-)gap associated
with the Hubbard bands: this is a distinguishing aspect of DMFT. It is crucial for an approximate
solver to reproduce this separation of energy scales in order to yield a correct description
of the Mott transition and phase diagram.


There are of course some differences between the three methods, on which we now comment.
First, we note that the IPT approximation has a somewhat larger quasiparticle bandwith.
This is because the transition point $U_{c2}$ is overestimated within IPT
(cf. Fig.~\ref{fig:phase}), so that a more fair comparison should perhaps be made at
fixed $U/U_{c2}$. It is true however, that the DSR method has a tendency to
underestimate the quasiparticle bandwith, and particularly at smaller values of
$U$. Accordingly, the Hubbard bands have a somewhat too large spectral weight,
but are correctly located in first approximation.
The detailed shape of the Hubbard bands is not very accurately known, in any case.
(The ED method involves a broadening of the delta-function peaks obtained by
diagonalizing the impurity hamiltonian with a limited number of effective orbitals,
so that the high-energy behaviour is not very accurate on the real axis. This is
also true, actually, of the more sophisticated numerical renormalization group).
We emphasize that, since the DSR method does not have the correct low-frequency
Fermi liquid behaviour, the quasiparticle bandwith should be interpreted as the
width of the central peak in $\rho_d(\omega)$ (while the quasiparticle weight $Z$
cannot be defined formally).

\begin{figure}[htbp]
\begin{center}
\vspace{0.6cm}
\includegraphics[width=7cm]{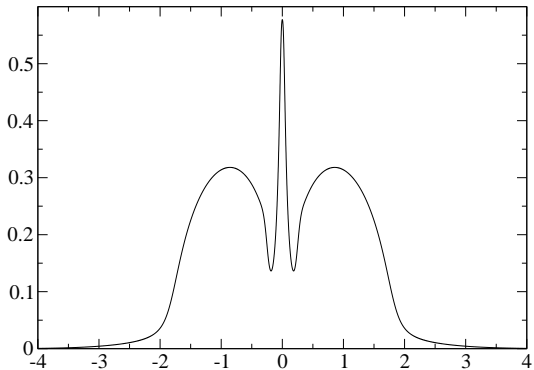}
\end{center}
\caption{$U$-NCA spectral function for $U=1.5$ at low temperature.}
\label{fig:unca}
\end{figure}

\begin{figure}[htbp]
\begin{center}
\vspace{0.6cm}
\includegraphics[width=7cm]{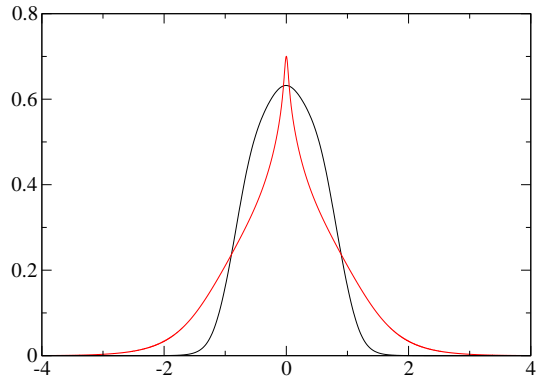}
\end{center}
\caption{Results for $U=0.1$ showing how $U$-NCA overshoots the Friedel's sum
rule. The Slave Rotor method is correctly converging towards the free density of states
(semi-circular)}
\label{fig:U_petit}
\end{figure}

In Fig~\ref{fig:unca} and Fig.~\ref{fig:U_petit}, we display the spectral functions
obtained by using the NCA method (extended at finite $U$ in the simplest manner).
It is clear that the $U$-NCA underestimates considerably the quasiparticle bandwith
(and thus yields a Mott transition at a rather low value of the coupling).
This is not very surprising, since this method is based on a strong coupling expansion
around the atomic limit, and one could think again of a comparison for fixed $U/U_c$.
More importantly, the $U$-NCA misses the important separation of energy
scales between the central peak and the Mott gap. As a result, it does not reproduce
correctly the phase diagram for the Mott transition within DMFT (in particular
regarding coexistence).
A related key observation is that the $U$-NCA does not have the correct weak-coupling
(small $U$) limit. To illustrate this point, we display in Fig.~\ref{fig:U_petit}
the $U$-NCA and DSR spectral functions for a tiny value of the interaction $U/D=0.1$:
the DSR result is shown to approach correctly the semi-circular shape of the non-interacting
density of states, while the $U$-NCA displays a characteristic inverted V-shape: in this regime the
violation of the Friedel sum rule becomes large and the negative lifetime pathology
is encountered within $U$-NCA. In contrast, the DSR yields a pinning of the
spectral density at a $U$-independent value and does not lead to a violation of
causality. It should be emphasized however that, even though it yields the exact $U=0$
density of states at $T=0$, the DSR method is not quantitatively very accurate in the weak-coupling regime.

\subsection{Double occupancy}
\label{sec:double}

Finally, we demonstrate that two-particle correlators in the charge sector
can also be reliably studied
with the DSR method, taking the fraction of doubly occupied sites
$\double \equiv
\left< n_\uparrow n_\downarrow \right>$ as an example.

Using the constraint (\ref{eq:contrainte}), we calculate the connected
charge susceptibility in the following manner:
\begin{eqnarray}
\nonumber
\chi_c(\tau) & \equiv & \left< \sum_\s \left(n_\s(\tau) -\frac{1}{2}\right) \sum_{\s'}
\left(n_{\s'}(0) -\frac{1}{2}\right) \right> \\
\label{eq:chicharge}
& = & \left< \hat{L}(\tau) \hat{L}(0) \right>
= \frac{1}{U^2} \left< \partial_\tau \theta(\tau) \partial_\tau \theta(0)
\right> \\
\nonumber
& = & \frac{2}{U^2} \left[ G_X(\tau) \partial_{\tau}^2 G_X(\tau)
+ U \delta(\tau) - (\partial_{\tau} G_X(\tau))^2 \right]
\end{eqnarray}
The double occupancy is obtained by taking the equal-time value:
\begin{equation}
\chi_c(\tau=0) = 2 \left< n_\uparrow n_\downarrow \right>  = 2 \double
\end{equation}

\begin{figure}[htbp]
\begin{center}
\vspace{0.6cm}
\includegraphics[width=7cm]{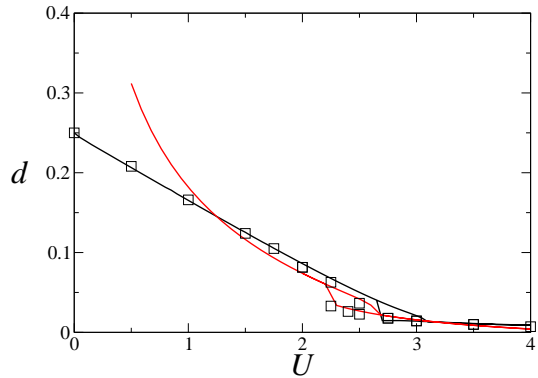}
\vspace{0.5cm}
\end{center}
\caption{Double occupancy at $\beta=80$ as a function of $U$. The ED result is
indicated with squares. DSR overshoots the exact result on the left,
whereas IPT is slightly displaced to the right.}
\label{fig:double}
\end{figure}

In Fig.~\ref{fig:double}, we plot the double occupancy obtained in that manner, in
comparison to the ED and IPT results. Within ED, $\double$ is calculated directly from the
charge correlator. Within IPT, $\chi_c(\tau)$ is not approximated very reliably
\cite{IPT_dynamique}, but the double occupancy can be accurately calculated by
taking a derivative of the internal energy with respect to $U$.

Fig.~\ref{fig:double} demonstrates that the DSR method is very accurate
in the Mott transition region and in the insulator. In particular, the hysteretic behaviour
is well reproduced. In the weak coupling regime however, the approximation deteriorates.
This issue actually depends on the quantity:
the physical Green's function has the correct limit
$U\rightarrow0$, as emphasized above, but this is not true for
the charge susceptibility (hence for $\double$).  The mathematical reason is
that the constraint (\ref{eq:contrainte}) is crucial for writing
(\ref{eq:chicharge}), but this constraint is only treated on average
within our method. This shows the inherent limitations of slave-boson
techniques for evaluating two-particle properties.
The frequency dependence
of two-particle correlators will be dealt with in a forthcoming publication.

\subsection{Multi-orbital effects}
\label{sec:multi-orbital}

In this section, we apply the DSR method to study the dependence of the Mott transition
on orbital degeneracy $N$.
We emphasize that there are at this stage very few numerical methods which can
reliably handle the multi-orbital case, specially when $N$ becomes large.
Multi-orbital extensions of IPT have been studied \cite{kajueter_multi}, but the
results are much less satisfactory than in the one-orbital case. The ED method is
severely limited by the exponential growth of the size of the Hilbert space.
The Mott transition has been studied in  the multi-orbital case using QMC, with
a recent study \cite{amadon_QMC} going up to $N=8$ (4 orbitals with spin).
Furthermore, we have recently
obtained \cite{mott_largeN} some analytical results on the values of the critical couplings
in the limit of large-$N$. Those, together with the QMC results,
can be used as a benchmark of the DSR approximation presented here.
As explained above, a value of $\cN_{N=2}\simeq 3$ was found to
describe best the single orbital case ($N=2$). Hence, upon increasing $N$, we choose
the parameter $\cN$ such that $\cN_N/\cN_{N=2}= N/2$.

\begin{figure}[htbp]
\begin{center}
\vspace{0.6cm}
\includegraphics[width=7cm]{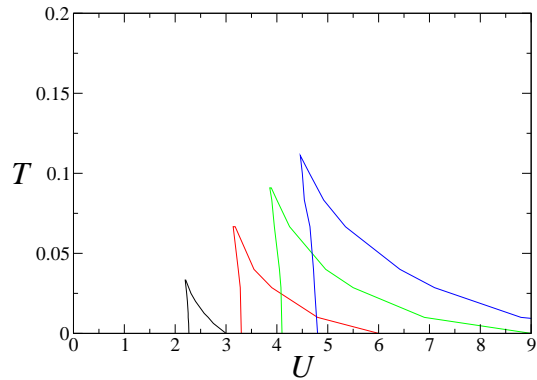}
\end{center}
\caption{Instability lines $U_{c1}(T)$,$U_{c2}(T)$, and coexistence region for
one, two, three and four orbitals ($N=2,4,6,8$)
as obtained by DMFT (DSR solver)}
\label{fig:lignes}
\end{figure}
In Fig.~\ref{fig:lignes}, we display the coexistence region in the $(U,T)$ plane
for increasing values of $N$, as obtained with DSR.
The values of the critical interactions grow with $N$.
A fit of the transition lines $U_{c1}(T)$ (where the insulator disappears)
and $U_{c2}(T)$ (where the metal disappears) yields:
\begin{eqnarray}
U_{c1}(N,T) & = & A_1(T) \sqrt{N} \\
U_{c2}(N,T) & = & A_2(T) N
\end{eqnarray}
These results are in good accordance with both the QMC data
\cite{amadon_QMC}, and the exact results established in \cite{mott_largeN}.
When increasing the number of orbitals, the coexistence region widens and
the critical temperature associated with the endpoint of the Mott transition line
also increases.

\begin{figure}[htbp]
\begin{center}
\vspace{0.6cm}
\includegraphics[width=7cm]{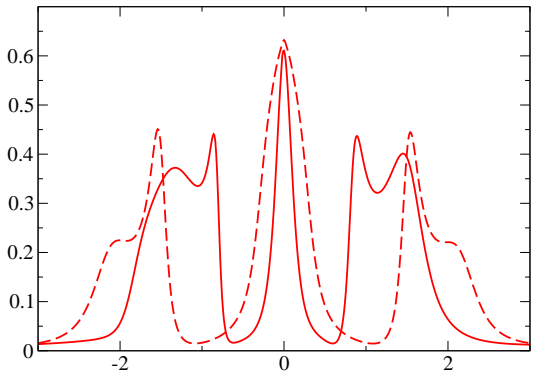}
\end{center}
\caption{Spectral function at $U=2$, $\beta=60$ and for one (full line)
and two orbitals (broken line)}
\label{fig:dos_N}
\end{figure}

We display on Fig.~\ref{fig:dos_N} the DSR result for the spectral function for $N=2$ and
$N=4$, at a fixed value of $U$ and $T$.
Two main effects should be noted. First, correlations effects in the metal become
weaker as $N$ is increased (for a fixed $U$), as clear e.g from the increase of the
quasiparticle bandwith.
This is due to increasing orbital fluctuations. Second, the Hubbard bands
also shift towards larger energies, an effect which can be
understood in the way atomic states broadens in the insulator
\cite{jarrell_multiorbital}.

To conclude this section, we have found that the DSR yields quite satisfactory
results when used in the DMFT context for multi-orbital models.
In the future, we plan to use this method in the context of realistic electronic structure
calculations combined with DMFT, in situations where ``numerically exact'' solvers
(e.g QMC) become prohibitively heavy.
There are however some limitations to the use of DSR (at least in the present version of the approach),
which are encountered when the occupancy is not close to $N/2$. We examine this issue
in the next section.

\subsection{Effects of doping}
\label{sec:doping}

Up to now, we studied the half-filled problem ({\it i.e.} containing exactly $N/2$ electrons),
with exact particle-hole symmetry.
In that case, $\epsilon_0\equiv\epsilon_d+U/2=0$ and the Lagrange multiplier $h$ enforcing
the constraint (\ref{eq:col4}) could be set to $h=0$.
In realistic cases, particle-hole symmetry will be broken (so that $\epsilon_0\neq 0$),
and we need to consider fillings different from $N/2$.
Hence, the Lagrange multiplier $h$ must be determined in order to fulfill
(\ref{eq:col4}), which we rewrite more explicitly as:
\begin{eqnarray}
n_f  & = &  \frac{1}{2} - \frac{2h}{\cN U} \\
\nonumber
& + &  \frac{1}{\cN U} \frac{1}{\beta}
\sum_n \frac{ -i \nu_n \left( e^{i \nu_n 0^+} +  e^{i \nu_n 0^-} \right)
}{ \nu_n^2/U + \lambda - 2ih\nu_n/U - \Sigma_X(i\nu_n) }
\end{eqnarray}
In this expression,
$n_f=\frac{1}{N}\sum_{\sigma} \langle f^{\dagger}_{\sigma} f_{\sigma}\rangle =
\frac{1}{N}\sum_{\sigma} \langle d^{\dagger}_{\sigma} d_{\sigma}\rangle$ is the average
occupancy per orbital flavor. (Note that the number of
auxiliary fermions and physical fermions coincide, as clear from
(\ref{eq:green})). $n_f$ is related to the d-level position (or chemical potential,
in the DMFT context) by:
\begin{equation}
n_f = \frac{1}{\beta} \sum_n \frac{e^{i \omega_n 0^+}}
{i\omega_n- \epsilon_0 + h - \Sigma_f(i\omega_n)}
\end{equation}

\begin{figure}[htbp]
\begin{center}
\vspace{0.6cm}
\includegraphics[width=7cm]{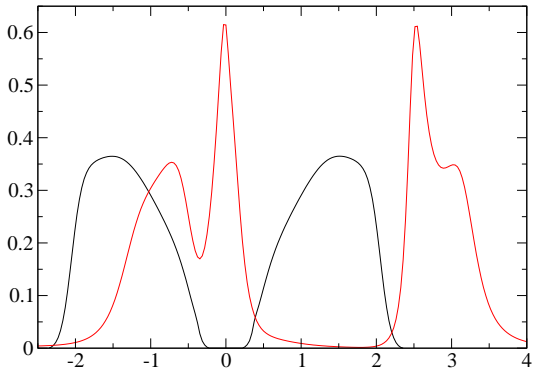}
\end{center}
\caption{Doping the insulator ($U=3$ and $\beta=60$) with $\eps=0,1$
(corresponding to $n_f=0.5,0.4$ respectively)}
\label{fig:dope_isolant}
\end{figure}

We display in Fig.~\ref{fig:dope_isolant} the spectral function obtained with the
DSR solver when doping the Mott insulator away from half-filling.
A critical value of the chemical potential ({\it i.e.} of $\epsilon_0$) is
required to enter the metallic state. The spectral function displays the
three expected features: a lower and upper Hubbard bands, as well as a quasiparticle
peak (which in this case where the doping is rather large, is located almost at the
top of the lower Hubbard band). However, it is immediately apparent from this
figure that the DSR method in the doped case overestimates the spectral weight of the
upper Hubbard band. This can be confirmed by a comparison to other solvers (e.g ED).
Note that the energy scale below which a causality violation appears within U-NCA becomes
rapidly large as the system is doped, while no such violation occurs within DSR.

The DSR method encouters severe limitations however as the total occupancy
becomes very different from $N/2$.
This is best understood by studying the dependence of the occupancy upon
$\epsilon_0$, in the atomic limit. As explained in
Appendix.~\ref{app:atomic}, the use of sigma-model variables $X$ (in the
large-$M$ limit) results in a poor description of the $n_f$ vs. $\epsilon_0$
dependence (``Coulomb staircase''). As a result, it is not possible to describe
the Mott transitions occurring in the multi-orbital model at integer fillings
different from $N/2$ using the DSR approximation. It should be emphasized
however that this pathology is only due to the approximation of the
$O(2)$ quantum rotor $e^{i\theta}$ by a $O(M)$ sigma-model field.
A perfect description of the Coulomb staircase is found for all fillings
when treating the constraint (\ref{eq:contrainte}) on average while
keeping a true quantum rotor \cite{hubbard_rotor}. Hence, it seems feasible to overcome
this problem and extend the practical use of the (dynamical) slave rotor
approach to all fillings. We intend to address this issue in a future work.

\section{Conclusion}

We conclude by summarizing the strong points as well as the limitations of the new
quantum impurity solver introduced in this paper, as well as possible extensions and
applications.

On the positive side, the DSR method provides an interpolating scheme between the
weak coupling and atomic limits (at half-filling). It is also free of some of the pathologies
encountered in the simplest finite-$U$ extensions of NCA (negative lifetimes at low temperature).
When applied in the context of DMFT, it is able to reproduce many of the qualitatively
important features associated with the Mott transition, such as coexisting insulating and metallic
solutions and the existence of two energy scales in the DMFT description of a correlated metal
(the quasi-particle coherence bandwith and the ``preformed gap'').
Hence the DSR solver is quite useful in the DMFT context, at a low computational cost, and
might be applicable to electronic structure calculations for systems close to half-filling when the orbital
degeneracy becomes large. To incorporate more realistic modelling, one can introduce different
energy levels for each correlated orbital, while the extension to non-symmetric
Coulomb interactions (such as the Hund's coupling) may require some additional work.

The DSR method does not reproduce Fermi-liquid behaviour at low energy however, which makes
it inadequate to address physical properties in the very low-energy regime (as is also the case with
NCA).
The main limitation however is encountered when departing from half-filling (i.e from
$N/2$ electrons in an $N$-fold degenerate orbital).
While the DSR approximation can be used at small dopings, it fails to reproduce the correct
atomic limit when the occupancy differs significantly from $N/2$ (and in particular cannot deal
with the Mott transition at other integer fillings in the multi-orbital case).
We would like to emphasize however that this results from extending
the slave rotor variable to a field with a large number of components. It is possible to
improve this feature of the DSR method by dealing directly with an $O(2)$ phase variable, which
does reproduce accurately the atomic limit even when the constraint is treated at the
mean-field level. We intend to address this issue in a future work. Another
possible direction is to examine systematic corrections beyond the saddle-point approximation
in the large $N,M$ expansion.

Finally, we would like to outline some other possible applications of the slave rotor representation
introduced in this paper (Sec.~\ref{sec:rotor}).
This representation is both physically natural and economical. In systems with strong Coulomb interactions,
the phase variable dual to the local charge is an important collective field. Promoting this single
field to the status of a slave particle avoids the redundancies of usual slave-boson
representations. In forthcoming publications, we intend to use this representation for:
i) constructing impurity solvers in the context of {\it extended} DMFT \cite{EDMFT},
in which the frequency-dependent charge correlation function must be calculated \cite{rotor_EDMFT}
ii) constructing mean-field theories of {\it lattice} models of correlated electrons (e.g the
Hubbard model) \cite{hubbard_rotor} and iii) dealing with quantum effects on the Coulomb blockade
in mesoscopic systems.

\acknowledgments{We are grateful to B.~Amadon and S.~Biermann for
sharing with us their QMC results on the multiorbital Hubbard model, and to
T.~Pruschke and P.~Lombardo for sharing with us their expertise of the NCA
method. We also acknowledge discussions with G.~Kotliar
(thanks to CNRS funding under contract PICS-1062) and with
H.~R.~Krishnamurthy (thanks to an IFCPAR contract No. 2404-1), as well as
with M. Devoret and D. Esteve.}

\appendix

\section{The Atomic limit}
\label{app:atomic}
\begin{figure}[htbp]
\begin{center}
\vspace{0.6cm}
\includegraphics[width=7cm]{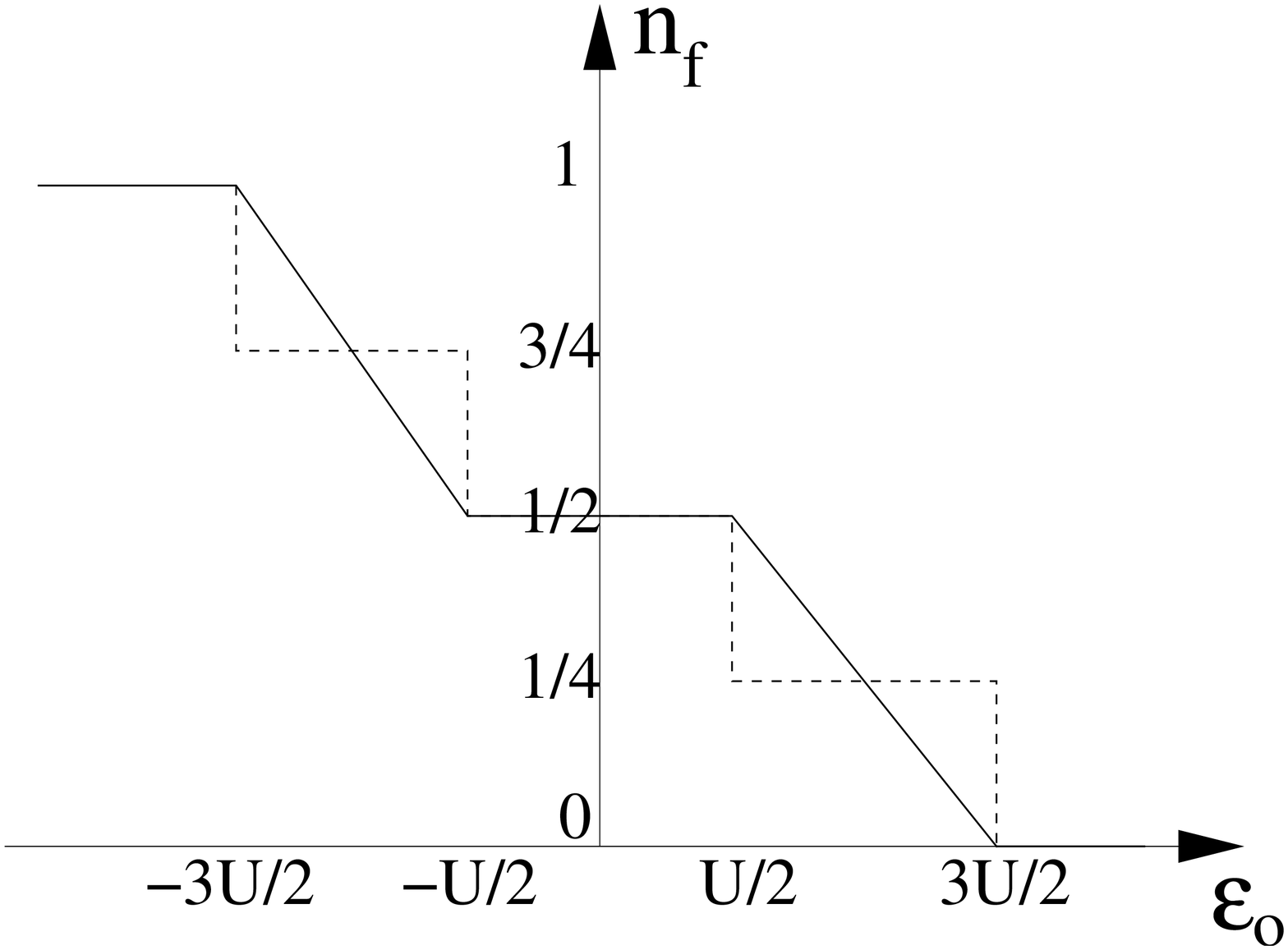}
\end{center}
\caption{Impurity occupancy in function of the $d$-level position in the rotor
description (full curve) and the exact result (dot curve), in the two-orbital case.}
\label{fig:staircase}
\end{figure}

In this appendix we prove the claim that the atomic limit of the model is exact
at half-filling and at zero temperature (at finite temperature, deviations from the exact result
are of order $\exp{(-\beta U)}$ and therefore negligeable for pratical purposes).
To do this, we first extract the values of the mean-field parameters $\lambda$
and $h$ from the saddle-point equations at zero temperature and
$\Delta(\tau)\equiv 0$.

\begin{eqnarray}
\label{eq:system_atomic}
1 &=& \int\! \frac{\mr{d}\nu}{2\pi}
\frac{1}{\frac{\nu^2}{U}+\lambda + \frac{2ih\nu}{U}}\\
\nonumber
\theta(h-\eps)
& = & \frac{1}{2} -\frac{2h}{\cN U} +\frac{4h}{\cN U^2} \int\! \frac{\mr{d}\nu}{2\pi}
\frac{\nu^2}{\left(\frac{\nu^2}{U}+\lambda\right)^2
+\left(\frac{2h\nu}{U}\right)^2}
\end{eqnarray}
Performing the integrals shows that $\lambda = (U^2-4 h^2)/(4U) $
and $\theta(h-\eps) = 1/2$, so that $h=\eps$. If $|\eps| > U/2$, the equations lead
actually to a solution with an empty or full valence that we show
on Fig.~\ref{fig:staircase}.

We can now compute the physical Green's function
$G_d(\tau) = G_f(\tau) G_X(-\tau)$ from the pseudo-propagators
$G_f(i\omega_n) = 1/(i\omega_n)$ and:
\begin{eqnarray}
G_X(i\nu_n) & = & \frac{1}{\nu_n^2/U+ (U/4 -\eps^2/U) - 2 i \eps \nu_n/U} \\
\nonumber
& = & \frac{-1}{i\nu_n + \eps - U/2} +  \frac{1}{i\nu_n + \eps + U/2}
\end{eqnarray}

Performing the convolution in imaginary frequency and taking the limit
$T=0$ leads to:
\begin{eqnarray}
\label{eq:convolution}
G_d(i\omega_n) & = & \frac{1}{\beta} \sum_{i\nu_n} G_X(i\nu_n)
G_f(i\omega_n+i\nu_n)\\
\label{eq:Gd_atomic}
& = & \frac{1/2}{i\omega_n-\eps+U/2}  + \frac{1/2}{i\omega_n-\eps-U/2}
\end{eqnarray}
Because $\eps = -\mu + U/2 $ this is the correct atomic limit of the
single-band model (at half-filling).
The result for the empty or full orbital is however not accurate, as shown in
figure \ref{fig:staircase}.
This discrepancy with the correct result (even for one orbital) finds its root
in the large $M$ treatment of the slave rotor $X$.

\section{Numerical solution of the real time equations}
\label{appendix1}
Here we show how the saddle-point equations (\ref{eq:col1}-\ref{eq:col4})
can be analytically continued along the real axis.
We start with $\Sigma_X(\tau) = -\cN \Delta(\tau) G_f(-\tau)$, which can
be first Fourier transformed into:
\begin{eqnarray}
\label{eq:continue}
\Sigma_X(i\nu_n) = & & \!\!\!\!\! \inte \;\Sigma_X(\tau) e^{i\nu_n \tau} \\
\nonumber
= -\cN \int \frac{\mr{d}\epsilon_1}{\pi}   & & \!\!\!\!\!\! G_f''(\epsilon_1)
 \int \frac{\mr{d}\epsilon_2}{\pi} \Delta''(\epsilon_2) \frac{ n_F(\epsilon_1) -
n_F(\epsilon_2)}{i\nu_n+\epsilon_1-\epsilon_2}
\end{eqnarray}
where we used the spectral representation
\begin{equation}
\label{eq:spectral}
G(z) = - \int \frac{\mr{d}\w}{\pi} \frac{G''(\epsilon)}{z-\epsilon}
\end{equation}
for each Green's function. The notations here are quite standard:
$G''(\epsilon) \equiv \mcal{I}m \, G(\epsilon+i0^+)$, $n_F(\epsilon)$ is the Fermi
factor, and $n_B(\epsilon)$ denotes the Bose factor.

It is then immediate to continue $i\nu_n \rightarrow \nu+i0^+$ in equation
(\ref{eq:continue}), and using again the spectral decomposition
(\ref{eq:spectral}), we derive an equation between retarded quantities:
\begin{eqnarray}
\label{eq:sxreal}
\Sigma_X(\nu) & = & -\cN \int \frac{\mr{d}\epsilon}{\pi}
G_f''(\epsilon) n_F(\epsilon) \Delta(\epsilon+\nu) \\
\nonumber
& & -\cN \int \frac{\mr{d}\epsilon}{\pi}
\Delta''(\epsilon) n_F(\epsilon) G_f(\epsilon-\nu)
\end{eqnarray}
A calculation along the same lines for the fermionic self-energy
$\Sigma_f(\tau) =  \Delta(\tau) G_X(\tau)$ leads to:
\begin{eqnarray}
\label{eq:sfreal}
\Sigma_f(\omega) & = & - \int \frac{\mr{d}\epsilon}{\pi}
G_X''(\epsilon) n_B(\epsilon) \Delta(\omega-\epsilon) \\
\nonumber
& & - \int \frac{\mr{d}\epsilon}{\pi}
\Delta''(\epsilon) n_F(\epsilon) G_X(\omega-\epsilon)
\end{eqnarray}

The numerical implementation is then straightforward because
(\ref{eq:sxreal}) and (\ref{eq:sfreal}) can each be expressed as
the convolution product of two quantities, so that they can be calculated
rapidly using FFT. The algorithm is looped back using the Dyson equations
(for real frequency):
\begin{eqnarray}
\label{eq:Gfreal}
G^{-1}_f(\omega) & = & \omega -\eps+h - \Sigma_f(\omega)\\
\label{eq:GXreal}
G^{-1}_X(\nu) & = & -\frac{\nu^2}{U}+\lambda +\frac{2 h\nu}{U} -\Sigma_X(\nu)
\end{eqnarray}

At each iteration, $\lambda$ and $h$ are determined using a bisection on the
equations (\ref{eq:col3}-\ref{eq:col4}), which can be properly expressed in
terms of retarded Green's functions:
\begin{eqnarray}
1 & = & \int \frac{\mr{d}\epsilon}{\pi} G_X''(\epsilon) n_B(\epsilon) \\
n_f & = & \frac{1}{2} - \frac{2h}{\cN U} - \frac{2}{\cN U}
\int \frac{\mr{d}\epsilon}{\pi} G_X''(\epsilon) \epsilon n_B(\epsilon)
\end{eqnarray}
where $n_f$, the average number of physical fermions, is:
\begin{equation}
n_f = G_f(\tau \myeq 0^-) = -  \int \frac{\mr{d}\epsilon}{\pi} G_f''(\epsilon)
n_F(\epsilon)
\end{equation}

We note here that solving these real-time integral equations can be quite
difficult deep in the Kondo regime of the Anderson model, or very close to the Mott
transition for the full DMFT equations.  The reason is that $G_f(\w)$
and $G_X(\w)$ develop low energy singularities (this is analytically shown in
the next appendix), that make the numerical resolution very unprecise if one
uses FFT. In that case, it is necessary to introduce a logarithmic mesh of
frequency (loosing the benefit of the FFT speed, but increasing the accuracy),
or to perform a Pad\'e extrapolation of the imaginary time solution.

\section{Friedel's sum rule}
\label{app:low}
We present here for completeness the derivation of the Slave Rotor Friedel's sum rule,
equation (\ref{eq:friedel}), at half-filling.
The idea, motivated by the numerical analysis as
well as theoretical arguments \cite{muller_hartmann,OP_AG_GK_AS}, is that the
pseudo-particles develop low frequency singularities at zero temperature:
\begin{eqnarray}
\label{eq:gfw}
G_f''(\w) & = & A_f |\w|^{-\alpha_f}\\
\label{eq:gXw}
G_X''(\w) & = & A_X |\w|^{-\alpha_X} \mr{Sign}(\w)
\end{eqnarray}
Using the spectral representation
\begin{equation}
\label{eq:repr}
G(\tau) = \int_0^{+\infty} \frac{\mr{d}\w}{\pi}\;
e^{-\w\tau} G''(\w)
\end{equation}
we deduce the long time behavior of the Green's functions (we denote by $\Gamma(z)$
the gamma function):
\begin{eqnarray}
\label{eq:gft}
G_f(\tau) &=& \frac{A_f \Gamma(1-\alpha_f)}{\pi} \frac{\mr{Sign}(\tau)}{|\tau|^{1-\alpha_f}}\\
\label{eq:gXt}
G_X(\tau) &=& \frac{A_X \Gamma(1-\alpha_X)}{\pi} \frac{1}{|\tau|^{1-\alpha_X}}
\end{eqnarray}
We have similarly
\begin{equation}
\Delta(\tau) = \frac{\Delta''(0)}{\pi\tau}
\end{equation}
 if one assumes a regular bath density of states at zero frequency. The previous expressions allow to extract the
long time behavior of the pseudo-self-energies (using the saddle-point equations
(\ref{eq:col1}-\ref{eq:col2})):
\begin{eqnarray}
\label{eq:sxw}
\Sigma_X(\tau) &=& \frac{\cN A_f \Gamma(1-\alpha_f)}{\pi^2} \frac{\Delta''(0)}{|\tau|^{2-\alpha_f}}\\
\label{eq:sfw}
\Sigma_f(\tau) &=& \frac{A_X \Gamma(1-\alpha_X)}{\pi^2} \frac{\Delta''(0)\;\mr{Sign}(\tau)}{|\tau|^{2-\alpha_X}}
\end{eqnarray}
The next step is to use (\ref{eq:repr}) the other way around to get
from (\ref{eq:sxw}-\ref{eq:sfw}) the $\w$-dependance of the self-energies:
\begin{eqnarray}
\Sigma_X''(\w) & = & \frac{\cN}{\pi} \frac{A_f \Delta''(0)}{1-\alpha_f} |\w|^{1-\alpha_f}\; \mr{Sign}(\w)\\
\Sigma_f''(\w) & = & \frac{1}{\pi} \frac{A_X \Delta''(0)}{1-\alpha_X} |\w|^{1-\alpha_X}
\end{eqnarray}
It is necessary at this point to calculate the real part of both self-energies.
This can be done using the Kramers-Kronig relation, but analyticity provides a
simpler route. Indeed, by noticing that $\Sigma(z)$ is an analytic function of
$z$ and must be uni-valuated above the real axis , we find:
\begin{eqnarray}
\nonumber
\Sigma_X(z) & = & \frac{\cN}{\pi} \frac{A_f \Delta''(0)}{1-\alpha_f}
\frac{e^{i(\alpha_f-1)\pi/2}}{\sin[(\alpha_f-1)\pi/2]} |z|^{1-\alpha_f}\\
\nonumber
\Sigma_f(z) & = & \frac{1}{\pi} \frac{A_X \Delta''(0)}{1-\alpha_X}
\frac{e^{i\alpha_X\pi/2}}{\sin[\alpha_X\pi/2]} |z|^{1-\alpha_X}
\end{eqnarray}

The same argument shows from (\ref{eq:gfw}-\ref{eq:gXw}) that:
\begin{eqnarray}
G_f(z) & = & A_f
\frac{e^{i(\alpha_f+1)\pi/2}}{\sin[(\alpha_f+1)\pi/2]} |z|^{-\alpha_f}\\
G_X(z) & = & A_X
\frac{e^{i\alpha_X\pi/2}}{\sin[\alpha_X\pi/2]} |z|^{-\alpha_X}
\end{eqnarray}

We can therefore collect the previous expressions, using Dyson's formula
for complex argument:
\begin{eqnarray}
\label{eq:Gfz}
G^{-1}_f(z) & = & z - \Sigma_f(z)\\
\label{eq:GXz}
G^{-1}_X(z) & = & -\frac{z^2}{U}+\lambda -\Sigma_X(z)
\end{eqnarray}
and this enables us to extract the leading exponents, as well as the product of
the amplitudes:
\begin{eqnarray}
\alpha_f &=& \frac{1}{\cN+1} \\
\alpha_X &=& \frac{\cN}{\cN+1} \\
A_f A_X &=& \frac{\pi}{\cN+1} \frac{1}{\Delta''(0)}
\sin^2\left(\frac{\pi}{2}\frac{\cN}{\cN+1} \right)
\end{eqnarray}

We finish by computing the long time behavior of the {\it physical} Green's function
$G_d(\tau) = G_f(\tau) G_X(-\tau)$ together with equations
(\ref{eq:gft}-\ref{eq:gXt}):
\begin{equation}
G_d(\tau) = \frac{\pi}{2(\cN+1)}\tan\left(\frac{\pi}{2} \frac{\cN}{\cN+1}\right)
\frac{1}{\pi \Delta''(0)} \frac{1}{\tau}
\end{equation}

This proves (\ref{eq:friedel}).
In principle, next leading order corrections can be computed by the same line of
arguments, although this is much more involved \cite{OP_AG_GK_AS}. Non Fermi Liquid
correlations in the physical Green's function would appear in this computation.

\end{document}